\documentclass[aps,prd,superscriptaddress,12pt,showpacs,notitlepage]{revtex4}
	\usepackage{graphicx}
	
\usepackage{epsf}
\usepackage{epsfig}

\newcommand{\be}{\begin{equation}}
\newcommand{\ee}{\end{equation}}
\newcommand{\bea}{\begin{eqnarray}}
\newcommand{\eea}{\end{eqnarray}}

\newcommand{\bb}{\bibitem}
 
\newcommand{\eqn}{\begin{eqnarray}}
\newcommand{\eqnx}{\end{eqnarray}}

\begin{document}
\title{The  vector BPS baby Skyrme model}
\author{C. Adam}
\affiliation{Departamento de F\'isica de Part\'iculas, Universidad de Santiago de Compostela and Instituto Galego de F\'isica de Altas Enerxias (IGFAE) E-15782 Santiago de Compostela, Spain}
\author{C. Naya}
\affiliation{Departamento de F\'isica de Part\'iculas, Universidad de Santiago de Compostela and Instituto Galego de F\'isica de Altas Enerxias (IGFAE) E-15782 Santiago de Compostela, Spain}
\author{J. Sanchez-Guillen}
\affiliation{Departamento de F\'isica de Part\'iculas, Universidad de Santiago de Compostela and Instituto Galego de F\'isica de Altas Enerxias (IGFAE) E-15782 Santiago de Compostela, Spain}
\author{A. Wereszczynski}
\affiliation{Institute of Physics,  Jagiellonian University,
Reymonta 4, Krak\'{o}w, Poland}

\pacs{11.30.Pb, 11.27.+d}

\begin{abstract}
We investigate the relation between the BPS baby Skyrme model and its vector meson formulation, where the baby Skyrme term is replaced by a coupling between the topological current $B_\mu$ and the vector meson field $\omega_\mu$. 
The vector model still possesses infinitely many symmetries leading to infinitely many conserved currents which stand behind its solvability. 
It turns out that the similarities and differences of the two models depend strongly on the specific form of the potential. 
 We find, for instance, that compactons (which exist in the BPS baby Skyrme model) disappear from the spectrum of solutions of the vector counterpart. 
 Specifically, for the vector model with the old baby Skyrme potential we find that it has compacton solutions only provided that a delta function source term effectively screening the topological charge is inserted at the compacton boundary. For the old baby Skyrme potential squared we find that the vector model supports exponentially localized solitons, like the BPS baby Skyrme model. These solitons, however, saturate a BPS bound which is a nonlinear function of the topological charge and, as a consequence, higher solitons are unstable w.r.t. decay into smaller ones, which is at variance with the more conventional situation (a linear BPS bound and stable solitons) in the BPS baby Skyrme model.    
\end{abstract}

\maketitle 
\section{Introduction}
In the low energy limit QCD becomes a strongly interacting quantum field theory where the primary quark and gluon fields are no longer proper degrees of freedom. Indeed, due to confinement they form colorless bound states, i.e., mesons and baryons (and perhaps glueballs). Although the complete description of these effective particles from first principles, that is by direct computations in QCD, is a very complicated and still unsolved problem, several effective theories have been proposed. One of the best known and successful approaches is the Skyrme framework \cite{skyrme}, in which pions are identified as the proper low energy degrees of freedom. Then, baryons appear as collective, topological excitations of the mesonic, matrix-valued field $U \in SU(2)$ with an identification between the baryon number and the topological charge $Q \in \pi_3(S^3)$. Unfortunately, the exact form of the effective action is unknown. Typically, one includes the standard sigma model part (kinetic term), the Skyrme part (the fourth derivative term which is required by the Derrick theorem) and a potential (chiral symmetry breaking term). A systematic derivation of the model from QCD has not been completed yet. Under additional assumptions (1-loop expansion, large $N_c$ limit and small derivative limit) one can derive Skyrme-like models, but with additional terms which seem to be rather problematic. They contain higher powers and higher orders of time derivatives that lead to problems with the (semi-classical) quantization and dynamical properties of solutions. Recently, a new Skyrme-type action has been proposed, the so-called {\it BPS Skyrme model} \cite{BPS-Sk1}, \cite{BPS-Sk2}. It consists of two parts: the baryon density current squared (the sextic term) and the potential. This model possesses many interesting mathematical properties, like infinitely many symmetries with infinitely many conserved charges, solutions saturating a BPS bound which are given by exact formulas for any value of the topological charge. In addition, it cures some of the phenomenological problems of the standard Skyrme model (strongly non-BPS solutions, large binding energies, lack of the incompressible fluid property). 
\\
A natural questions one may ask, once a Skyrme type model is specified, is how the properties of the solitons (baryons) are changed by interactions with the lightest fields i.e., the $U(1)$ Maxwell field and the vector mesons $\omega_\mu$. Specifically, the interaction between the Skyrme field and the vector mesons is given by the coupling of the vector field with the topological current density $\omega_\mu B^\mu$ \cite{vec1}-\cite{vec3}. Such a coupling is known to effectively induce a term proportional to the square of the baryon current. Because of the high nonlinearity of the vector Skyrme model we do not know much about its solitonic solutions. Usually one is left with numerical solutions \cite{Sutcliffe vector}, and a detailed understanding of the relation between solitons in the Skyrme and the vector Skyrme model is still rather incomplete.   
\\
In this work we want to analytically investigate this issue in $(2+1)$ dimensions using the low dimensional version of the BPS Skyrme model \cite{old} i.e., {\it the BPS baby Skyrme model} \cite{GP}-\cite{Sp1}. In this lower-dimensional case, the term induced by the vector meson (the topological current squared) is, at the same time, the usual Skyrme term which is of fourth power in derivatives. In particular, we will show for which potentials both models (the BPS baby and its vector version) have qualitatively similar solitonic solutions, and for which potentials the properties of solutions are rather distinct.   
\\
An additional, related aim of this paper is to shed further light on possible applications of  the BPS Skyrme model to baryon physics. The BPS Skyrme model with the standard potential (i.e., the pion mass term) may be viewed as a chiral theory where the pions are very (infinitely) massive in the sense that the normal linear pion excitations are completely suppressed. On the other hand, pions are the lightest effective particles. Hence, the inclusion of the vector mesons with finite mass and a standard kinetic term in the BPS Skyrme model in some sense reverses the typical mass hierarchy and, therefore, it would be surprising  if it  lead to qualitatively similar solitons as in the original BPS model. Here, we show that in 2+1 dimensions the solitons in the baby BPS model with the standard mass term for the "pions" (the so-called old baby Skyrme potential), and, therefore, with infinitely heavy "baby" pions in the sense just explained, are in fact completely different from the solitons in its vector version (with finite mass "baby" vector mesons). These findings should be contrasted with an analogous investigation for the full  baby Skyrme model \cite{baby vector} with the "old" baby Skyrme potential, where the authors find that the solitons in the original baby Skyrme model and its vector version behave rather similarly. 
\\
Finally, let us mention that one way to pass from the BPS baby Skyrme model to the vector model is by replacing the local integration kernel in the Skyrme term with a nonlocal integral kernel of the Yukawa or Coulomb type. That is to say, instead of the Skyrme term
$$
L_{Sk} = B_\mu^2 (x)= \int d^3 y B_\mu (x) g^{\mu\nu} \delta^3 (x-y) B_\nu (y)
$$
we have the nonlocal term
$$
L_{vec} = \int d^3 y B_\mu (x) K^{\mu\nu} (x-y) B_\nu (y)
$$
where $K^{\mu\nu} (x-y)$ is the integral kernel of the Yukawa or Coulomb type which results from integrating out the (massive or massless) vector mesons. The same nonlocal interaction induced by a Yukawa or Coulomb integral kernel has been investigated for the nonlinear Schroedinger equation in \cite{HaZa}.   
\section{The $\omega$-vector model}
It is well-known known that the coupling with a vector field offers an alternative way to stabilize baby skyrmions \cite{baby vector}. Then, instead of the standard (2+1) dimensional Skyrme term one introduces a coupling between the topological field and vector mesons by means of the topological current $B^{\mu}$. Concretely, the Lagrange density reads 
\begin{equation}
\mathcal{L}=\frac{m^2}{2} (\partial_\mu \vec{\phi} )^2-\mu^2 V(\vec{\phi}) - \frac{1}{4} (\partial_\mu \omega_\nu - \partial_\nu \omega_\mu)^2 +\frac{1}{2} M^2 \omega_\mu^2 +\lambda' \omega_\mu B^\mu,
\end{equation}
where the topological current is
\begin{equation}
B^{\mu}=-\frac{1}{8\pi} \epsilon^{\mu \alpha \beta} \vec{\phi} \cdot  (\partial_{\alpha} \vec{\phi} \times \partial_\beta \vec{\phi} ) ,
\end{equation}
and $\vec{\phi}=(\phi^1,\phi^2,\phi^3)$ is a three component unit vector field. It introduces a nontrivial topology into the model since, for static configurations, it may be viewed as a map from $R^2 \cup \{ \infty \} = S^2$ (obtained by a compactification of the base space via adding a point at infinity, because $\vec{\phi} \rightarrow \phi_{vac}$ when $\vec x \rightarrow \infty$) into the target space $S^2$.  Therefore, we will call this field topological, in contrast to the mesonic fields $\omega_{\mu}$. The potential $V$ is a positive definite function which depends only on the topological field, and whose particular form influences rather strongly the properties of baby skyrmions \cite{holom}, \cite{new}. Here, we restrict ourselves to potentials which leave some of the global $O(3)$ symmetry unbroken, i.e., which depend only on the third component of the topological field $V=V(\phi^3).$ More concretely, we will analyze a family of one-vacuum potentials which provides a natural generalization of the old baby potential \cite{old},
\begin{equation}
V=\left( \frac{1-\phi_3}{2} \right)^\alpha 
\end{equation}
where $\alpha \geq 1$. 
\\
The extreme (or BPS) version of the model is defined by the same assumption as in the case of the usual baby Skyrme model, that is $m=0$, i.e., when the standard kinetic term for the topological field is absent
\begin{equation}
\mathcal{L}=-\mu^2 V(\phi_3) - \frac{1}{4} (\partial_\mu \omega_\nu - \partial_\nu \omega_\mu)^2 +\frac{1}{2} M^2 \omega_\mu^2 +\lambda' \omega_\mu B^\mu
\end{equation}
Observe that we leave the mesonic part of the model untouched, so it is by no means obvious whether all the nice features of the BPS baby Skyrme model will survive also in its vector meson formulation. 
\\
For reasons of convenience we prefer to deal with the $CP^1$ version of the model, that is, we apply the stereographic projection
\be
\vec{\phi}=\frac{1}{1+|u|^2} \left( u+\bar{u}, -i ( u-\bar{u}),
|u|^2-1 \right) .
\ee 
Then, 
\begin{equation}
\mathcal{L}=-\mu^2 V(|u|^2) -  \frac{1}{4} (\partial_\mu \omega_\nu - \partial_\nu \omega_\mu)^2 +\frac{1}{2} M^2 \omega_\mu^2+ i\lambda \omega_\mu \epsilon^{\mu \alpha \beta} \frac{u_\alpha \bar{u}_\beta}{(1+|u|^2)^2}
\end{equation}
with potential 
\begin{equation}
V= \left( \frac{|u|^2}{1+|u|^2} \right)^\alpha 
\end{equation}
The pertinent field equations are
\begin{equation}
\partial_\mu^2 \omega_\nu +M^2\omega_\nu + i\lambda \epsilon_{\nu} ^{\;\; \alpha \beta} \frac{u_\alpha \bar{u}_\beta}{(1+|u|^2)^2}=0,
\end{equation}
\begin{equation}
i\lambda \epsilon^{\alpha \mu \beta} \partial_\mu \omega_\alpha \bar{u}_\beta +\alpha \mu^2  \left( \frac{|u|^2}{1+|u|^2} \right)^{\alpha -1}  \bar{u}=0 ,
\end{equation}
and their static versions are
\begin{equation}
\partial_i^2\omega-M^2\omega= i\lambda \frac{\nabla_r u \nabla_\phi \bar{u}-\nabla_r \bar{u} \nabla_\phi u}{(1+|u|^2)^2},
\end{equation}
\begin{equation}
i\lambda(\nabla_r \omega \nabla_\phi \bar{u}-\nabla_r \bar{u} \nabla_\phi \omega ) +\alpha \mu^2\left( \frac{|u|^2}{1+|u|^2} \right)^{\alpha -1}  \bar{u}=0
\end{equation}
where we have used that $B_i =0$, $\omega_i=0$ and $\omega \equiv \omega_0$.
An obvious ansatz which is compatible with these equations is the axially symmetric one
\begin{equation}
u=f(r) e^{in\phi}, \;\;\; \omega=\omega(r),
\end{equation}
where $n$ is an integer number which coincides with the winding number of the topological field. Further, the relevant topologically nontrivial boundary conditions are
\begin{equation}
f(r=0)=\infty, \;\;\;\; f(r=R)=0
\end{equation}
where, depending on the potential, $R$ may be finite (for compactons) or infinite (for exponentially or power-like localized baby skyrmions).
Additionally, the regularity of the equation (finiteness of the energy) for the meson field requires
\begin{equation}
\omega'_r (r=0)=0, \;\;\; \omega (r=R)=0.
\end{equation}
Finally, one usually imposes the first derivative continuity (or regularity for non-compacton case) condition
\begin{equation}
f' (r=R) =0, \;\;\;\; \omega'(r=R)=0.
\end{equation}
As we will see below, however, in some cases this condition cannot be satisfied. This results in an important qualitative difference between solutions of the BPS Skyrme model with the old baby potential and its $\omega$ meson version. 
\subsection{Massless case}
Let us start with the simplest case of massless $\omega$ mesons. The limit of massless vector mesons is not directly related to the induction of the Skyrme term, but it is simpler and some features of the solutions will be useful later on. The massless vector meson field may also be interpreted as an abelian $U(1)$ field which here is coupled to the BPS baby model by the topological current only. Such a coupling, alternative to the minimal one, does not require an unbroken $U(1)$ symmetry for the topological field $\vec{\phi}$. Hence, in principle, one may consider potentials which completely break the global $SO(3)$ symmetry. Notice that a similar non-minimal coupling appears as part of the (3+1) dimensional Skyrme model coupled properly to the Maxwell field \cite{witten ED}. Let us also emphasize that the analogy with the electromagnetic interaction allows us to conclude immediately that static solutions have to obey one of the two following conditions. Either the energy of the static solution is infinite, because the electrostatic energy in two space dimensions is infinite (has a logarithmic IR divergence). Or the total electric charge 
and, therefore, the topological charge (which is equal to the electric charge in our case) is, in fact, zero, despite our topologically nontrivial ansatz. This second possibility implies that the corresponding static solution is really a solution in the presence of an inhomogeneous source term forced upon us by the consistency of the solution, such that the source effectively screens the topological charge. We shall find both possibilities in the sequel.
The static field equations are
\begin{equation}
\frac{1}{r} \partial_r (r\omega'_r)= \frac{2n\lambda}{ r} \frac{ff'_r}{(1+f^2)^2} \label{massless r1}
\end{equation}
\begin{equation}
f \left[ \frac{1}{r} \omega'_r +\frac{ \alpha \mu^2}{n\lambda}  \left( \frac{f^2}{1+f^2} \right)^{\alpha -1}  \right]=0 .\label{massless r2}
\end{equation}
Now we introduce the new variable $x=r^2/2$ and get
\begin{equation}
\partial_x (2x\omega'_x)= 2n\lambda \frac{ff'_x}{(1+f^2)^2}
\;\;\;\; \Rightarrow \;\;\;\;
\partial_x \left[ 2x\omega'_x + n\lambda \frac{1}{1+f^2} +C \right]=0 \label{massless1}
\end{equation}
where $C$ is an integration constant
and
\begin{equation}
f \left[ \omega'_x +\frac{\alpha \mu^2}{n\lambda} \left( \frac{f^2}{1+f^2} \right)^{\alpha -1} \right]=0. \label{massless2}
\end{equation}
The last equation allows to express $\omega'_x$ in terms of the profile function $f$. Hence, after inserting it into (\ref{massless1}) we get a simple algebraic formula for $f$
\begin{equation}
(1+f^2) \left( \frac{f^2}{1+f^2} \right)^{\alpha -1} = \frac{n^2 \lambda^2}{2\alpha \mu^2} \frac{1}{x} 
\equiv \frac{X_0}{\alpha x}
\label{f massless}
\end{equation}
where we have used the boundary condition $f(x=0)=\infty$ to fix the integration constant $C=0$. Then, the solution should be inserted back into (\ref{massless2}) to obtain $\omega$. 
\subsubsection{Compacton - $\alpha =1$}
As a first example let us consider the old baby potential i.e., $\alpha=1$. Then one finds  
\begin{equation}
f(x)=\left\{ 
\begin{array}{cc}
\sqrt{\frac{X_0}{x}-1} & x \leq X_0 \\
 & \\
 0 & x > X_0
\end{array} \right. \label{comp1}
\end{equation}
\begin{equation}
\omega(x)=\left\{ 
\begin{array}{cc}
\frac{n\lambda}{2} \left(1-\frac{x}{X_0} \right) & x \leq X_0 \\
 & \\
 0 & x > X_0
\end{array} \right.
\end{equation}
where 
\begin{equation}
X_0=\frac{n^2 \lambda^2}{2 \mu^2} \label{comp2}
\end{equation}
One can see that this solution is only of $\mathcal{C}$-class (i.e. a continuous function without continuous derivatives), and, therefore, may introduce a source term at the compacton boundary.   
\vspace*{0.2cm}

\noindent {\bf Proposition 1:} Any compacton solution of the massless e.o.m. (\ref{massless1}), (\ref{massless2}) is a solution with a Dirac delta source located at the boundary of the compacton which screens the total baryon charge inside the compacton. 
\\
{\bf Proof:}
We start by assuming a compacton solution with a finite compacton radius $R$
\begin{equation}
f(r)=\left\{ 
\begin{array}{cc}
\tilde{f}(x)& x \leq X \\
 & \\
 0 & x > X
\end{array} \right.
\end{equation}
\begin{equation}
\omega(x)=\left\{ 
\begin{array}{cc}
\tilde{\omega}(x) & x \leq X \\
 & \\
 0 & x > X
\end{array} \right.
\end{equation}
where $\tilde{f}, \tilde{\omega}$ solve (\ref{massless1}), (\ref{massless2}) inside the ball with the radius $X$. Specifically, for the massless vector model with the old baby potential they are given by (\ref{comp1}), (\ref{comp2}) and $X=X_0$. Now, consider the equation for the meson field inside the compact ball (we use that $C=0$)
\begin{equation}
\partial_x \left( 2x\omega'_x \right) = - n\lambda \partial_x \left(  \frac{1}{1+\tilde{f}^2} \right) .\label{source1}
\end{equation}
Here, the right hand side, that is the topological charge density, may be treated as a source for the mesonic field
$$\rho (x) = - n  \partial_x \left(  \frac{1}{1+\tilde{f}^2} \right) .$$ 
We easily find the total charge inside the compacton ball 
\begin{equation}
Q_{inside}=2\pi \int_0^X dx \rho (x) = -2\pi n \int_0^X dx \partial_x \left(  \frac{1}{1+\tilde{f}^2} \right) = -2\pi n  \left.\frac{1}{1+\tilde{f}^2} \right|_0^X = -2\pi n .
\end{equation}
However, when we look at equation (\ref{source1}) at the boundary (or equivalently in the whole space) then an additional current appears. Namely, we define a function $H(x)$ 
\begin{equation}
H(x)=2x\omega_x'+n \lambda \frac{1}{1+f^2}
\end{equation}
whose derivative is equal to the static equation for the mesonic field (\ref{massless1}). Due to the equations of motion, it holds that $H(x)=0$ inside the ball. On the other hand, for the vacuum values $f=0$ and $\omega = 0$ which the fields take outside the compacton we get $H(x)=\lambda$. Therefore, the function $H$ computed for compacton solutions is proportional to a Heaviside theta function
\begin{equation}
H(x)|_{on \; shell} =n \lambda \theta (X-x) .
\end{equation}
Obviously, its derivative is proportional to the Dirac delta, and a source term at the compacton boundary emerges. Thus, finally, the compacton solution is a solution to the following equation 
\begin{equation}
\partial_x \left( 2x\omega'_x \right) = - n\lambda \partial_x \left(  \frac{1}{1+\tilde{f}^2} \right) + n\lambda \delta (X-x) . \label{source2}
\end{equation}
The topological charge stored in the compact ball is screened by the source term added at the boundary of the compacton. Hence, the total charge of the solution is zero which results in the finiteness of the total energy.
\subsubsection{Non-compact solutions - $\alpha >1 $}
Now, from (\ref{f massless}) one sees that the profile function tends to 0 at spatial infinity in a power-like fashion. Indeed, solving (\ref{f massless}) for large $x$ we get as the leading term
\begin{equation}
f(x) = \left(\frac{X_0}{x} \right)^{\frac{1}{2(\alpha -1)}}+...
\end{equation}
On the other hand, from (\ref{massless1}) 
\begin{equation}
\omega'_x = -\frac{n\lambda}{2 x}+... \quad \Rightarrow \quad \omega = -\frac{n\lambda}{2} \ln x + ...
\end{equation} 
and it is precisely this logarithmic behaviour which will give rise to the IR divergent energy.
To see it, let us consider the static energy integral
\begin{eqnarray}
E = 2\pi \int_0^\infty rdr\left[ \mu^2V(f) - \frac{1}{2}\omega^2_r -\frac{2\lambda n}{r} \frac{\omega ff'_r}{(1+f^2)^2} \right]  &  & \\ =  2\pi  \int_0^\infty dx \left[ \mu^2V(f) - x \omega'^2_x -2\lambda n\frac{\omega ff'_x}{(1+f^2)^2} \right] .
\end{eqnarray}
As always for a "Coulomb" energy, the second and third term combine into a total derivative plus a term which would normally give the Coulomb energy. The difference is that here the total derivative is IR divergent. Indeed, using the (non-dynamical) field equation (\ref{massless1}) (with $C=0$) for $x\omega_x'$ we get
\begin{equation}
E = 2\pi \int_0^\infty dx \left[ \mu^2 V(f) + \frac{n\lambda}{2}\frac{d}{dx}\left( \frac{\omega}{1+f^2}\right) + \frac{n\lambda}{2} \omega \frac{d}{dx} \frac{1}{1+f^2 } \right]
\end{equation}
where the third term is the normal Coulomb term, whereas the second, total derivative term is IR divergent in our case due to the $\omega \sim \ln x$ behaviour of $\omega$ for large $x$.

As an exactly solvable example we may consider the case $\alpha=2$. Then  
\begin{equation}
f(x)=
\sqrt{\frac{X_0}{x}}, 
\end{equation}
\begin{equation}
\omega (x)=
 - \frac{n\lambda }{2} \ln (x+X_0) .
\end{equation}
\\
Undoubtedly, the massless vector meson version of the BPS baby Skyrme model differs significantly from its standard BPS baby counterpart. They are two rather different theories and their topological solutions do not have much in common. The main difference is that the vector BPS baby model in the massless limit does not possess finite energy solitonic solutions of the sourceless equations of motion. 
\subsection{Massive case}
Now we allow for a nonzero mass of the vector mesons. Then, only one equation of motion is modified. Namely, after the insertion of the ansatz into the static equations we find
\begin{equation}
 n\lambda \partial_x \left[  \frac{1}{1+f^2} \right]=M^2 \omega - \partial_x (2x \omega'_x) \label{massive1}
\end{equation}
while the second equation obtained by variation with respect to the complex scalar remains unchanged  
\begin{equation}
f \left[ \omega'_x +\frac{\alpha \mu^2}{n\lambda} \left( \frac{f^2}{1+f^2} \right)^{\alpha -1} \right] =0 . \label{massive2}
\end{equation}
\subsubsection{Compacton - $\alpha=1$}
In this case, the last equation can be completely integrated with the following result
\begin{equation}
\omega=\frac{\mu^2}{n\lambda} (X-x)
\end{equation}
for $x \leq X$ and $\omega=0$ for $x >X$, where at the moment $X$ is an arbitrary constant - the size of the compacton. Now, we insert it into the first equation  
\begin{equation} 
 \partial_x \left[  \frac{1}{1+f^2} \right]= -\frac{\mu^2M^2}{n^2\lambda^2} x + \frac{2\mu^2}{n^2\lambda^2} \left( 1+M^2X \right)
\end{equation}
which can be easily solved. After imposing the boundary conditions we get
\begin{equation}
\frac{1}{1+f^2(x)}=\left\{ 
\begin{array}{cc}
-\frac{M^2}{2X_0}x^2 + \frac{x}{X_0} \left( 1+M^2 X  \right) & x \leq X \\
 & \\
 1 & x > X
\end{array} \right.
\end{equation}
\begin{equation}
\omega(x)=\left\{ 
\begin{array}{cc}
\frac{n \lambda}{2} \left(\frac{X}{X_0}-\frac{x}{X_0} \right) & x \leq X \\
 & \\
 0 & x > X
\end{array} \right.
\end{equation}
where 
\begin{equation}
X=\frac{1}{M^2} (-1 +\sqrt{1+2M^2X_0} )
\end{equation}
In the limit of small mass $M  \rightarrow 0$ we find that $X \rightarrow X_0$ and we re-derive the former result. One can also check that
\begin{equation}
X \leq X_0
\end{equation}
and the size of the compacton tends to zero as $M \rightarrow \infty$. So, the massive vector meson squeezes the compact baby skyrmion.  
\\
Finally, one can again prove that our solution solves the field equations with the same source term as in the massless case. This implies that, again, the topological charge is screened and that the solution is not a genuine topological soliton, although in the massive vector meson case there is no general physical principle which would enforce this result.  And, indeed, we shall see in the next section that genuine finite energy solitons may exist for massive vector mesons. 
\subsubsection{Exponentially localized solution - $\alpha = 2$ case}
\noindent In order to find solutions for other values of the parameter $\alpha$ we differentiate (\ref{massive1}) and then insert (\ref{massive2}). This gives a second order ODE for the profile function
\begin{equation}
\partial_x^2 \left( \frac{1}{1+f^2} - \frac{x}{X_0} \left( \frac{f^2}{1+f^2} \right)^{\alpha -1} \right)= - \frac{M^2}{2X_0} \left( \frac{f^2}{1+f^2} \right)^{\alpha -1} . \label{eq-mass}
\end{equation}
This equation is quite complicated and, in general,  no analytical solution can be found. However, there is one exception, namely the case with $\alpha=2$. Then, we find
\begin{equation}
\partial_x^2 \left( \frac{1}{1+f^2} - \frac{x}{X_0}  \frac{f^2}{1+f^2} \right)= - \frac{M^2}{2X_0}  \frac{f^2}{1+f^2}, \;\;\; X_0(\alpha=2)=\frac{n^2\lambda^2}{4\mu^2} 
\end{equation}
This equation can be brought to a linear form after the substitution 
$
g=\frac{f^2}{1+f^2}$ and $ x=\frac{\sqrt{2X_0}}{M} y$
\begin{equation}
\partial_y^2 \left[ \left( 1+\beta y)g \right) \right]= g, \;\;\;\beta=\frac{\sqrt{2}}{\sqrt{X_0}M} 
\end{equation}
or 
\begin{equation}
g'' (1+\beta y) +2\beta g'-g=0
\end{equation}
with the boundary conditions $g(0)=1$ and $g(\infty)=0$ (here we already anticipate that the solutions are not compactons). Then the solution is 
\begin{equation}
g(y)= \frac{1}{K_1(2/\beta) } \frac{K_1 \left(  \frac{2}{\beta} \sqrt{1+\beta y} \right)}{\sqrt{1+\beta y}}  
\end{equation}
where $K_1$ is the modified Bessel function of the second type.  Hence, $g(y)$ is a smooth function which goes to zero exponentially at spatial infinity. The same happens for the derivative of $\omega$
\begin{equation}
\omega_y= -\sqrt{2} \frac{ \mu }{M} g = -\sqrt{2} \frac{ \mu }{M} \frac{1}{K_1(2/\beta) } \frac{K_1 \left(  \frac{2}{\beta} \sqrt{1+\beta y} \right)}{\sqrt{1+\beta y}} .
\end{equation}
Now, we want to describe the energy of this configuration. In the field $g(x)$ and variable $x$, the energy can be written as
\begin{eqnarray} 
E &=& 2 \pi \int^\infty_0  dx \left[ \mu^2 g^2 -  x \omega_x^2 - \frac{1}{2} M^2 \omega^2 - n \lambda  \omega g_x \right] \nonumber \\
&=& 2 \pi \int^\infty_0  dx \left[ \mu^2 g^2 -  \frac{1}{2} n \lambda  \omega g_x - 
\frac{d}{dx} \left( x\omega \omega_x \right) \right] \label{E1}
\end{eqnarray}
where we used a partial integration and the (non-dynamical) equation
\begin{equation} \label{massive1a}
-\frac{d}{dx} \left( 2x\omega_x \right) + M^2 \omega = -n\lambda g_x
\end{equation}
for $\omega$ to eliminate the non-dynamical field $\omega$. The resulting term $-(1/2) n\lambda \omega g_x$ (where $\omega$ must be expressed by its solution) corresponds to the Coulomb self energy in the massless vector meson case, so we might call it the "Yukawa self energy" in the case of a massive vector meson. The corresponding Yukawa self-energy density is, in fact,  positive semi-definite, because $\omega (x) \ge 0$ and $g_x (x) \le 0$. Finally, the total derivative term gives zero for massive vector mesons because $x\omega \omega_x$ is zero both at $x=0$ and at $x=\infty$. 

\begin{figure}[h]
 \begin{center}
  {\includegraphics[width=0.55\textwidth]{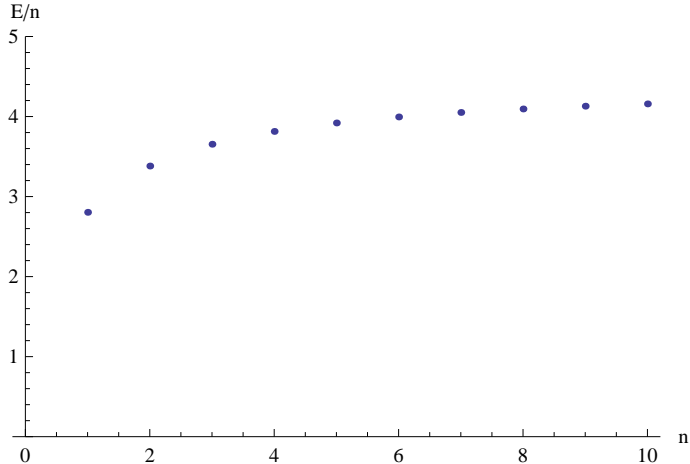}} 
  \caption{Energy per topological charge $n$, as a function of $n$, for $\alpha =2$, and 
for the parameter values $M = \mu  = \lambda =1$. The sublinear corrections reducing the energies of small solitons are clearly visible.}
  \label{Energy_n}
 \end{center}
\end{figure}

The energy is, in fact, of the BPS type, that is, it may be expressed as a total derivative with the help of the static field equations. Indeed, using
\begin{equation}
\omega_x = -\frac{2\mu^2}{n\lambda} g
\end{equation}
(which follows from Eq. (\ref{massive2})) the energy may be re-expressed like
\begin{equation} 
E = 2\pi \frac{n^2 \lambda^2}{4\mu^2 } \int_0^\infty dx (\omega_x^2 + \omega \omega_{xx}) =
-\frac{\pi}{2} \left( \frac{n\lambda}{\mu}\right)^2 \omega (0) \omega_x (0).
\end{equation}   
Further, $g(0)=1$ leads to $\omega_x (0)=-(2\mu^2/n\lambda )$, whereas from Eq. (\ref{massive1a}), which in terms of the independent variable $y$ may be written like
\begin{equation}
g_y (1+\beta y) + \beta g = -\frac{M}{\sqrt{2}\mu} \omega
\end{equation}
we get 
\begin{equation}
\omega (0) = \sqrt{2}\frac{\mu}{M} \left( -\frac{\beta}{2} + 
\frac{K_0 \left(\frac{2}{\beta} \right) + K_2 \left(\frac{2}{\beta} \right)
}{2K_1 \left(\frac{2}{\beta} \right)} \right) 
\end{equation}
where we used that the $y$ derivative of $g(y)$ is
\begin{equation}
g_y = -\frac{\beta}{2} \frac{ K_1 \left( \frac{2 \sqrt{1+\beta y}}{\beta} \right) }{ (1+\beta y)^\frac{3}{2}   
K_1 \left(\frac{2}{\beta} \right) } - \frac{ K_0 \left( \frac{2 \sqrt{1+\beta y}}{\beta} \right) +
K_2 \left( \frac{2 \sqrt{1+\beta y}}{\beta} \right) }{ 2(1 + \beta y) K_1 \left( \frac{2}{\beta} \right) } .
\end{equation}
Putting everything together, we get the following energy expression
\begin{eqnarray}
E &=& \sqrt{2}\pi \frac{n\lambda \mu}{M} \left( -\frac{\beta}{2} + 
\frac{K_0 \left(\frac{2}{\beta} \right) + K_2 \left(\frac{2}{\beta} \right)
}{2K_1 \left(\frac{2}{\beta} \right)} \right) \\
&\simeq & \sqrt{2}\pi \frac{n\lambda \mu}{M} \left( 1-\frac{1}{4}\beta + \frac{3}{32} \beta^2 + \ldots \right) 
\end{eqnarray}
where
\begin{equation}
\beta = \frac{\sqrt{2}}{\sqrt{X_0} M} = \frac{2 \mu}{n\lambda M}
\end{equation}
and we performed a small $\beta$ (i.e., large $n$) expansion in the last line. We find that the BPS energies are not exactly linear in the topological charge $n$. Instead, they have sub-leading corrections in inverse powers of $n$, and the leading correction (proportional to $\beta$, i.e., $n^{-1}$) has a negative sign. This implies that the masses of small solitons of charge $n=1$ are {\em less} than $1/n$ times the masses of large solitons with charge $n$, and the large solitons are therefore {\em energetically unstable} w.r.t. decay into smaller ones, see Fig. \ref{Energy_n}.
It might still be possible that there exist higher charge solitons without rotational symmetry with lesser energy, which would then be less unstable or even stable.
All this is quite different from the case of the standard BPS baby skyrmions, where already the rotationally symmetric solitons saturate a bound exactly linear in the topological charge and are, therefore, stable. We remark that only for the case $\alpha = 2$ we were able to express the on-shell energy as a total derivative, which implies that the spherically symmetric soliton solutions saturate a BPS bound. We were not able to find a similar result for other potentials.   

\subsubsection{Exponentially localized solution - $\alpha \in (1,2)$ case}

\begin{figure}[h]
 \begin{center}
  {\includegraphics[width=0.55\textwidth]{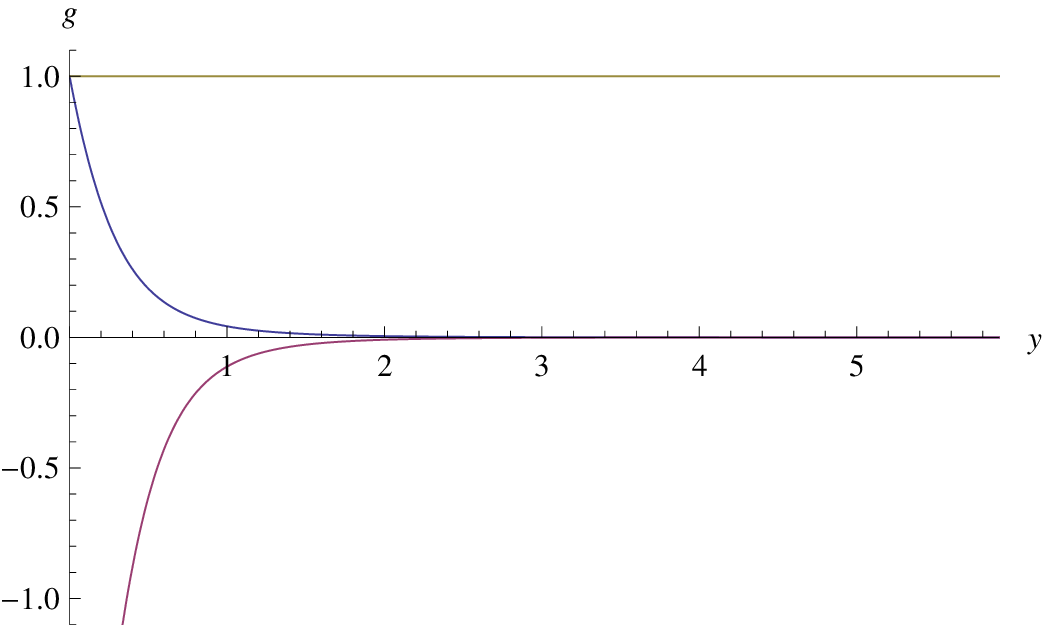}} \\
 {\includegraphics[width=0.55\textwidth]{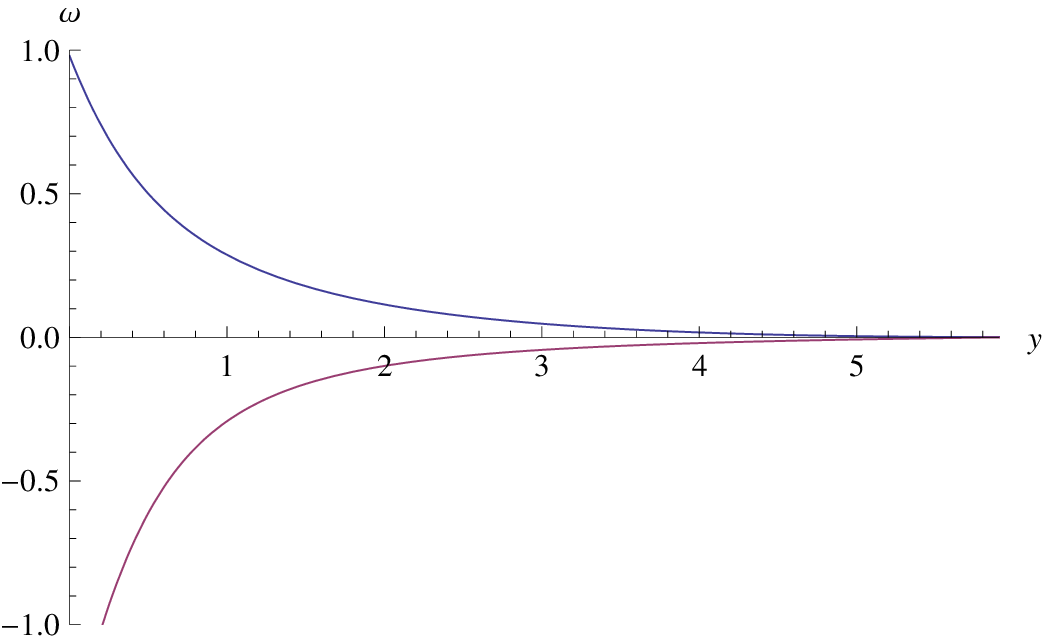}} \\
  \caption{Solutions $g, \omega$ and their derivatives for $\alpha=3/2$, 
for the parameter values $M = \mu = n = \lambda =1$.}
  \label{sol_a23}
 \end{center}
\end{figure}

For other values of $\alpha \in (1,2)$ the power expansion at the vacuum value $f=0$ or $g=0$ in (\ref{eq-mass}) also leads to exponentially localized solutions. This follows from the fact that for $g \rightarrow 0$ the field equation
\begin{equation}
\partial_y^2 \left[ g + \beta y g^{\alpha-1}  \right]=g^{\alpha-1}
\end{equation}
can be approximated by 
\begin{equation}
 \beta \partial_y^2 \left[ y g^{\alpha-1}  \right]=g^{\alpha-1} ,
\end{equation}
as $g < g^{\alpha-1}$ when $0<(\alpha -1) <1$. Then, the asymptotic solution, again, has an exponential tail
 \begin{equation}
 g \sim \frac{e^{-\frac{2}{\alpha-1} \sqrt{\frac{y}{\beta} } }}{ (\beta y)^{\frac{1}{2(\alpha-1)} } }
 \end{equation}
Obviously, the vector meson field as well as the energy density are exponentially localized, as well. 
\\
An example of such a solution is given by $\alpha=3/2$. 
In this case we solve numerically the equations by shooting from the center. In this way we find the solutions presented in Fig. \ref{sol_a23}, where the initial values (i.e, values of the free integration constants) for this solution are
\begin{equation}
g'(0)=-3.201, \qquad \qquad \qquad \omega(0)=0.92043
\end{equation}
\subsubsection{Power-like localized solutions - $\alpha >2$}
\begin{figure}[h]
 \begin{center}
 {\includegraphics[width=0.55\textwidth]{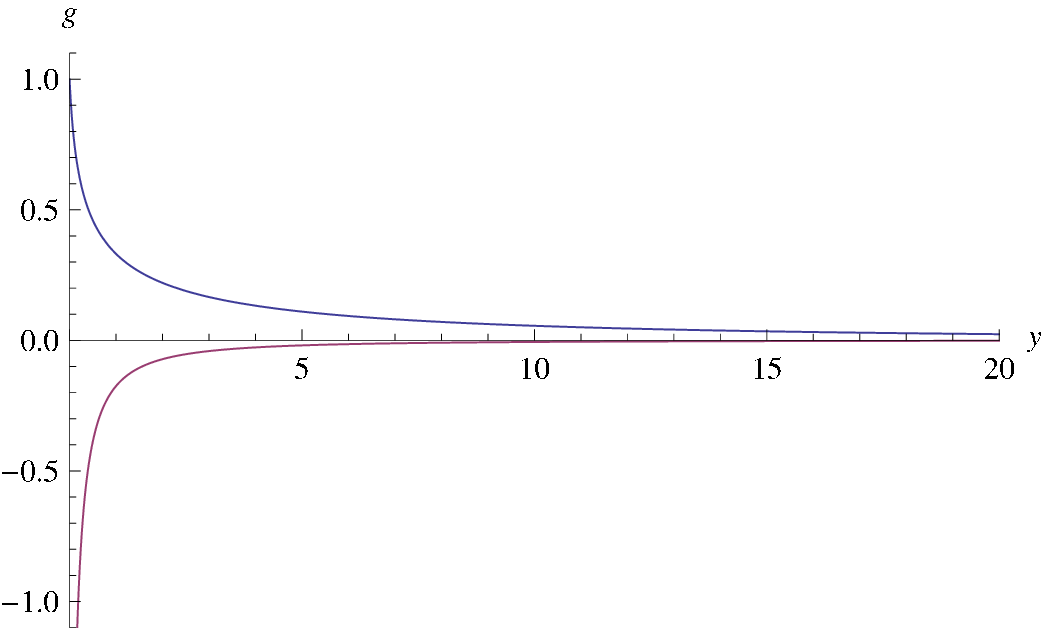}} \\
{\includegraphics[width=0.55\textwidth]{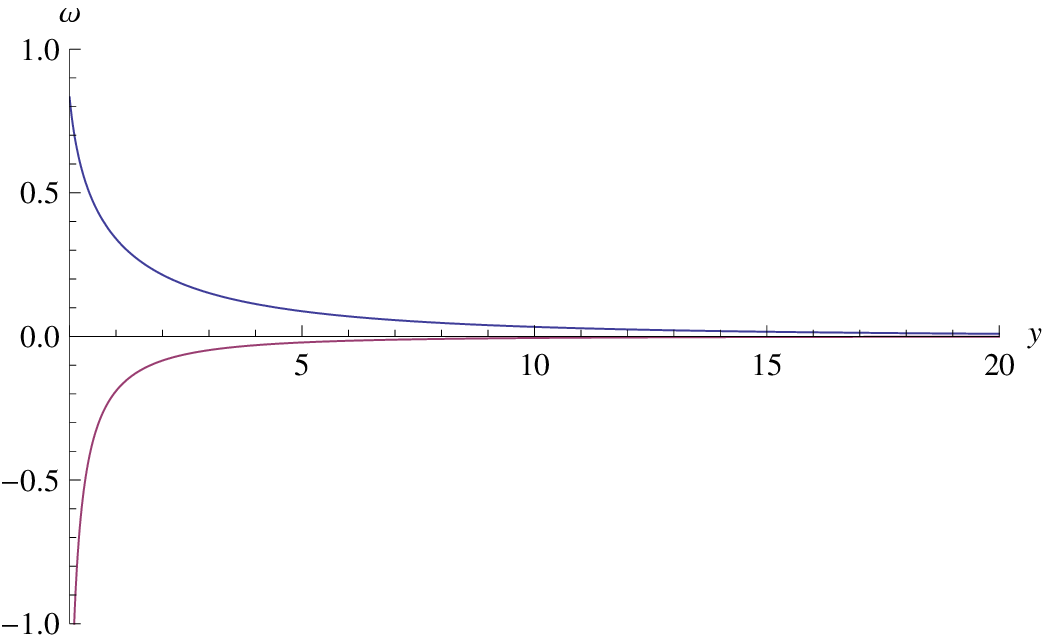}} \\
  \caption{Solutions $g, \omega$ and their derivatives for $\alpha=3$,
  for the parameter values $M = \mu = n = \lambda =1$.}
  \label{sol_a3}
 \end{center}
\end{figure}

 It is easy to find that for $\alpha >2$ we get a power-like tail for the profile function 
\begin{equation}
f(x) \sim \left( \frac{1}{x} \right)^{\frac{1}{\alpha-2}}
\end{equation}
which is exactly equal to the behavior of the profile function in the original BPS baby Skyrme model \cite{restr-bS}.  Then, one can obtain the asymptotic behaviour of the meson function
\begin{equation}
\omega'_x \sim \left( \frac{1}{x} \right)^{2\frac{\alpha-1}{\alpha-2}}  \;\;\; \Rightarrow \;\;\; \omega \sim  \left( \frac{1}{x} \right)^{2\frac{\alpha-1}{\alpha-2}-1}
\end{equation}
This behaviour is confirmed by the numerical analysis. For example, for $\alpha=3$ and for the parameter values $M=\mu =\lambda =n= 1$ we find for the free integration constants for a soliton solution 
\begin{equation}
g'(0)=-3.94465, \qquad \qquad \qquad \omega(0)=0.8327,
\end{equation}
which gives the solutions presented in Fig. \ref{sol_a3}. 
\subsection{Infinite mass limit}
We have demonstrated that there are no topologically nontrivial solutions to the vector version of the BPS baby Skyrme model with the old baby potential, unless some source terms are added. Thus, the vector BPS baby Skyrme model with the old baby Skyrme potential is {\it qualitatively} very different from the original BPS baby Skyrme model. 
\\
From a phenomenological point of view this is not surprising. The BPS baby Skyrme model
is based on the  assumption of very (infinitely) heavy "baby" pions in the sense of a complete suppression of linear "pion" excitations.
Physically, however, pions are the lightest mesons. Hence, in order to maintain the same physical hierarchy of masses, in this limit also other mesons should be infinitely heavy. 
\\
On the other hand, in the vector BPS baby Skyrme model the vector mesons possess a well defined finite mass $M$ while we still have infinitely heavy baby pions (no sigma term for the topological field). Thus, both models refer to a distinct mass hierarchy for the underlying particle d.o.f. Obviously, the limit represented by the vector model is not a physical one. Therefore, the fact that we do not reproduce results from the BPS baby Skyrme model in its vector version is a nice observation. Otherwise, the BPS model could be viewed (by its vector version) as a model with a completely unphysical assumption for the masses of the baby mesonic particles.
\\
In fact, there is a limit which brings the vector baby BPS model back to the BPS Skyrme model i.e, all baby mesonic particles have infinite masses. We simply assume that $M \rightarrow \infty$ i.e., we omit the kinetic part for the $\omega$-mesons in the Lagrangian. Of course, we arrive at a model in which the vector mesons are no longer dynamical d.o.f but rather a kind of Lagrange multipliers   
\begin{equation}
\mathcal{L}=-V(\phi_3) +\frac{1}{2} M^2 \omega_\mu^2 +\lambda' \omega_\mu B^\mu .
\end{equation}
Now, the field equations for the $\omega$-mesons are
\begin{equation}
\omega_\mu = -\frac{\lambda'}{M^2} B_\mu
\end{equation}
and can be used to eliminate these fields from the Lagrangian. Finally we re-derive the BPS baby Skyrme model 
\begin{equation}
\mathcal{L}=-\mu^2V(\phi_3) - \frac{1}{2} \frac{\lambda'^2}{M^2} B_\mu^2 .
\end{equation}
\section{Integrability and conservation laws}
The last issue which needs to be investigated is how the interaction with the vector mesons influences the integrability properties. In fact, we will show that the model is integrable in the sense of generalized integrability \cite{gen-int}. The first family of infinitely many conserved currents is given by the expression
 \begin{equation}
 j_\mu^G = i G(u\bar{u}) \left( \bar{u} \bar{\pi}_\mu - u \pi_\mu \right)
\end{equation}
where $G$ is an arbitrary function of the modulus of the complex scalar field and $\pi_\mu$ and $\bar{\pi}_\mu$ denote the canonical momenta 
\begin{equation}
\bar{\pi}_{\mu}=\frac{\partial \mathcal{L}}{\partial \bar{u}^{ \mu}} =i\lambda \omega_\alpha \epsilon^{\alpha \beta \mu} \frac{u_\beta}{(1+|u|^2)^2}
\end{equation}
\begin{equation}
\pi_{\mu}=\frac{\partial \mathcal{L}}{\partial u^{\mu}} =i\lambda \omega_\alpha \epsilon^{\alpha \mu \beta} \frac{\bar{u}_\beta}{(1+|u|^2)^2} .
\end{equation}
The equations of motion can be written as 
\begin{equation}
\partial_\mu \pi^\mu = -\mu^2V'\bar{u}-2i\lambda \bar{u} \omega_\alpha \epsilon^{\alpha \mu \beta} \frac{u_\mu \bar{u}_\beta}{(1+|u|^2)^3}
\end{equation}
and
\begin{equation}
\partial_\mu \bar{\pi}^\mu = -\mu^2V' u-2i\lambda u \omega_\alpha \epsilon^{\alpha \mu \beta} \frac{u_\mu \bar{u}_\beta}{(1+|u|^2)^3} .
\end{equation}
Then,
\begin{equation}
\partial_\mu j_G^\mu = iG'  \left( \bar{u} u_\mu + u \bar{u}_\mu \right) \left( \bar{u} \bar{\pi}_\mu - u \pi_\mu \right)+iG  \left( \bar{u} \partial^\mu \bar{\pi}_\mu - u \partial^\mu \pi_\mu \right)+  i G\left( \bar{u}^\mu \bar{\pi}_\mu - u^\mu \pi_\mu \right) .
\end{equation}
Using the identities which follow form the form of the momenta
\begin{equation}
\pi_\mu \bar{u}^\mu=0, \;\;\;\; \bar{\pi}_\mu u^\mu=0
\end{equation}
and the field equations we get 
\begin{equation}
\partial_\mu j^\mu_G =  i (G' u\bar{u}+ G)\left( \bar{u}^\mu \bar{\pi}_\mu - u^\mu \pi_\mu \right) .
\end{equation}
But
\begin{equation}
\bar{u}^\mu \bar{\pi}_\mu = i\lambda \omega_\alpha \epsilon^{\alpha \beta \mu} \frac{u_\beta \bar{u}_\mu}{(1+|u|^2)^2}=u^\mu \pi_\mu
\end{equation}
and, therefore
\begin{equation}
\partial_\mu j^\mu_G = 0 .
\end{equation}
Notice that we have used only the field equation for the complex scalar which is the same in the massless and massive $\omega$-meson models. Hence, the currents are conserved in both cases.  
\\
For the massless model we can construct another infinite family of conserved currents. Now they have the form
 \begin{equation}
 j_\mu^H=H(u\bar{u}) F_{\mu \nu} ( \bar{u}u^\nu + u \bar{u}^\nu) = H(u\bar{u}) F_{\mu \nu} \partial^\nu ( \bar{u}u) 
\end{equation}
where $H$ is an arbitrary function of the modulus of the complex scalar. We find
\begin{equation}
\partial^\mu j_\mu^H = H F_{\mu \nu} \partial^\mu \partial^\nu ( \bar{u}u) + H' F_{\mu \nu}  \partial^\mu ( \bar{u}u)  \partial^\nu ( \bar{u}u) +H (\partial^\mu F_{\mu \nu}) \partial^\nu ( \bar{u}u) .
\end{equation}
The first two terms vanish due to the contraction of the antisymmetric tensor $F_{\mu \nu}$ with two symmetric ones. Using the field equation for the mesons we get for the third term
\begin{equation}
\partial^\mu j_\mu^H = -i\lambda H \epsilon^{\nu \alpha \beta} \frac{u_\alpha \bar{u}_\beta}{(1+|u|^2)^2}  ( \bar{u}u^\nu + u \bar{u}^\nu)=0.
\end{equation}

\section{Summary and conclusions}

\begin{figure}[h]
\begin{center}
\includegraphics[width=0.8 \textwidth]{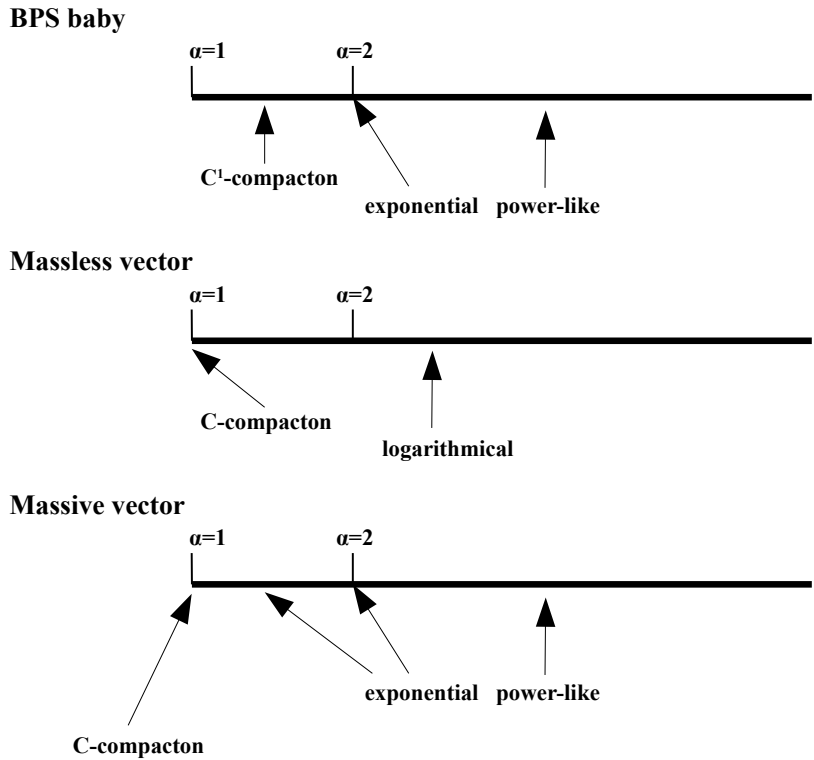}
\caption{Comparison of types of solutions in the BPS baby Skyrme model with its massless and massive vector counterparts. }
\label{3-models}
\end{center}
\end{figure}

The vector BPS baby model is integrable (in the sense of generalized integrability \cite{gen-int}) and to a far extent solvable for arbitrary values of the topological charge carried by a solitonic solution. So, in these aspects it reflects very well properties of the BPS baby Skyrme model. 
\\
In the case of the old baby Skyrme potential, which may be viewed as a mass term for the Skyrme field, the BPS and vector models lead to quite different solutions. Namely, the vector model (massive as well as massless) requires a delta source term located at the boundary of the compacton, which equivalently can be stated as the appearance of a $\mathcal{C}$-compacton, while in the original BPS baby model we get a $\mathcal{C}^1$-compacton, that is, a solution to the sourceless equations of motion. So, strictly speaking, the vector BPS baby model with this potential does not support sourceless solitonic solution. 
This should be contrasted with the analogous situation for the full baby Skyrme model with the old baby Skyrme potential, where the soliton solutions of the original model and its vector meson version are very similar, see \cite{baby vector}.
As we pointed out already, the different behaviour in our case is a welcome result from the point of view of phenomenological applications to baryon physics. If the same happens in (3+1) dimensions (for the vector BPS Skyrme model) then the BPS Skyrme model would preserve the mass hierarchy of the mesonic fields in QCD. Otherwise, the BPS Skyrme model (infinitely heavy pion approximation) could be described by a model with finite massive vector mesons, and this is not acceptable from a phenomenological point of view as the pions are the lightest mesonic degrees of freedom. Undoubtedly, this should be checked in the future.   
\\
For other potentials, where the "pion" potential cannot be interpreted as a mass term, we found that in some cases the soliton solutions in the original BPS baby Skyrme model and the massive vector model are more similar. Concretely,  for $\alpha \geq 2$ we observe exactly the same localization of solutions for the same values of the potential parameter $\alpha$ (exponential for $\alpha =2$,  or power-like for $\alpha >2$),  i.e., the same approach to the vacuum. For the specific case $\alpha =2$ we were able to find exact solutions and deduced from these solutions that, still, there remain some important differences. The solitons of the vector model for $\alpha =2$ saturate a BPS bound, like the solutions of the BPS baby Skyrme model, but the nature of this BPS bound in the vector model is quite different. It is, in fact, a nonlinear function of the topological charge which constitutes a rather interesting result by itself
(more recent results on nontrivial BPS bounds can be found, e.g. in \cite{Baz1}, \cite{gaugeBPSbaby}). As a consequence, a soliton with topological charge $n$ is heavier than $n$ solitons with charge one and is, therefore, energetically unstable against a decay into smaller solitons. This is completely different from the situation in the original BPS baby Skyrme model, where the solitons saturate a BPS bound which is linear in the topological charge and are, therefore, stable.  
The standard, linear BPS bound is, in fact, quite important for applications to hadronic or nuclear physics \cite{BPS-Sk1}, \cite{BPS-Sk2}, \cite{BoMa}, \cite{Sut1}. 
\\
For $\alpha \in (1,2)$, the compactons of the original BPS baby model transform into exponentially localized solitons with the tail becoming steeper and steeper as one approaches $\alpha=1$.  
It follows that the compactons, which exist in the original BPS baby Skyrme model for some potentials, completely disappear in the vector version. They either become solutions of the field equation with a source term ($\alpha=1$), or they receive exponential tails ($\alpha \in (1,2)$). The disappearance of the compactons may perhaps be understood from the fact that the vector meson field possesses the standard kinetic  and mass terms, which might prevent the solutions from becoming compactons. However, in the gauged BPS baby Skyrme model the addition of the standard kinetic term for the abelian gauge field does not lead to any problems with compactons \cite{gaugeBPSbaby}. Therefore, the detailed form of the coupling of the additional (standard) field to the BPS baby model also plays a significant role. 
\\
In the case of the massless model, all potentials with $\alpha >1$ generate large distance solutions with a logarithmically IR divergent energy, and this result can be understood easily by a comparison with the electrostatic Coulomb energy in two space dimensions.  
\\
We summarize the types of topological solutions (baby skyrmions) we found in the three models in Fig. \ref{3-models}.

\vspace*{0.3cm}

{\centerline {\bf Acknowledgement}}

\vspace*{0.2cm}

The authors acknowledge financial support from the Ministry of Education, Culture and Sports, Spain (grant FPA2008-01177), the Xunta de Galicia (grant INCITE09.296.035PR and Conselleria de Educacion), the Spanish Consolider-Ingenio 2010 Programme CPAN (CSD2007-00042), and FEDER. CN thanks the Spanish Ministery of Education, Culture and Sports for financial support (grant FPU AP2010-5772). Further, AW was supported by polish NCN grant 2011/01/B/ST2/00464.

\end{document}